\documentclass[12pt,amsmath,amssymb]{revtex4}
\usepackage{graphics}
\usepackage{amsfonts}
\usepackage{amsmath}
\usepackage{epsfig}
\usepackage{epsf}
\usepackage{graphicx}
\usepackage{dcolumn}
\usepackage{bm}
\usepackage{color}
\usepackage{array}
\usepackage{amssymb}
\usepackage{breqn}
\usepackage{multirow}

\newcommand {\sla}[1]{ #1 \!\!\!/}

\begin{document}

\title{Mass shifts of $^3P_J$ heavy quarkonia due to the effect of two-gluon annihilation}
\author{
Hui-Yun Cao, Hai-Qing Zhou \protect\footnotemark[1] \protect\footnotetext[1]{E-mail: zhouhq@seu.edu.cn} \\
School of Physics,
Southeast University, NanJing 211189, China}
\date{\today}

\begin{abstract}
In this work, we calculate the nonrelativistic asymptotic behavior of the amplitudes of $q\overline{q}\rightarrow 2g \rightarrow q\overline{q}$ in the leading order of $\alpha_s$ (LO-$\alpha_s$) with $q\overline{q}$  in the $^3P_J$ channels. In the practical calculation we take the momenta of quarks and antiquarks on-shell and expand the amplitudes on the three-momentum of the quarks and antiquarks to order 6 and get three nonzero terms. The imaginary parts of the first term and the second term are the old. The real parts of the results have IR divergence. When applying the results to the heavy quarkonia, the corresponding amplitude of $q\overline{q}\rightarrow1g\rightarrow q\overline{q}$ with $q\overline{q}$ in the color octet $^3S_1$ channel is considered to absorb the IR divergence in a unitary way in the leading order of $v$ (LO-$v$). The finial results can be used to estimate the mass shifts of the $^3P_J$ heavy quarkonia due to the effect of two-gluon annihilation. The numerical estimation shows that the contributions to the mass shifts of $\chi_{c0,c1,c2}$ are about $1.23\sim1.58$ MeV, $1.57\sim1.86$ MeV and $5.92\sim5.45$ MeV when taking $\alpha_s\approx 0.25\sim0.35$.
\end{abstract}

\maketitle

\section{\label{sec-1}Introduction}

The energy spectrum of an elemental system is a basic question after the breakthrough of quantum mechanism. Currently, the energy spectrum of hadrons are still an unsolved problem in QCD due to the complex nonperturbative property. Many phenomenological models have been used to studies the energy spectrum of hadrons in the quark level such as the quark model\cite{quark-model}, QCD sum rules\cite{QCD-sumrule}, Dyson-Schwinger equation and Bethe-Salpeter equation \cite{BS-eq},  etc. In these calculations the annihilation effect whose imaginary and real parts correspond to the decay width and the mass shift is usually neglected. For heavy quarkonia, their inclusive decays can be well described by the effective theory nonrelativistic QCD (NRQCD)\cite{NRQCD}. In NRQCD, the imaginary part of the coefficients of four fermions interactions are matched from the imaginary parts of the on-shell scattering amplitudes $q\overline{q}\rightarrow 1g$ or $2g$ or $3g \rightarrow q\overline{q}$ or the decay widths of $q\overline{q}\rightarrow 1g$ or $2g$ or $3g$ in perturbative QCD order by order. In previous paper \cite{CaoHuiYun2018}, we calculated the real parts of these coefficients in the leading order of $\alpha_s$ (LO-$\alpha_s$) in the $^1S_0$ channel with the momenta of quarks and antiquarks off-shell and find the results are gauge invariant, while the similar calculation can not be directly extended to the $^3P_J$ channels due to the gauge invariance.  In this paper, we follow the idea of NRQCD and calculate the amplitudes of $q\overline{q}\rightarrow2g\rightarrow q\overline{q}$  in the $^3P_J$ channels and  $q\overline{q}\rightarrow1g\rightarrow q\overline{q}$  in the $^3S_1$ channel with $q\overline{q}$ in color single and color octet states, respectively.

We organize the paper as follows. In Sec. II we give an introduction on the basic formula, in Sec. III we describe our calculation and present the analytic results for the coefficients to order 6 after the nonrelativistic expansion,  in Sev. IV we estimate the effects to the mass shifts numerically and discuss the interesting properties of the results.

\section{\label{sec-2} Basic formula}
\begin{figure}[htbp]
\center{\epsfxsize 3.0 truein\epsfbox{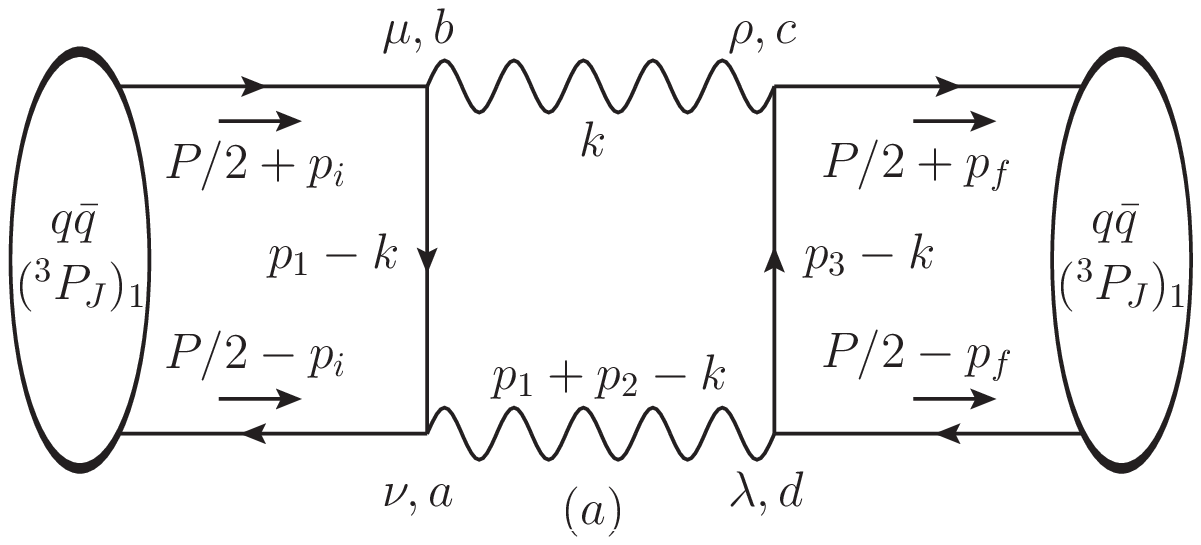}\epsfxsize 3.0 truein\epsfbox{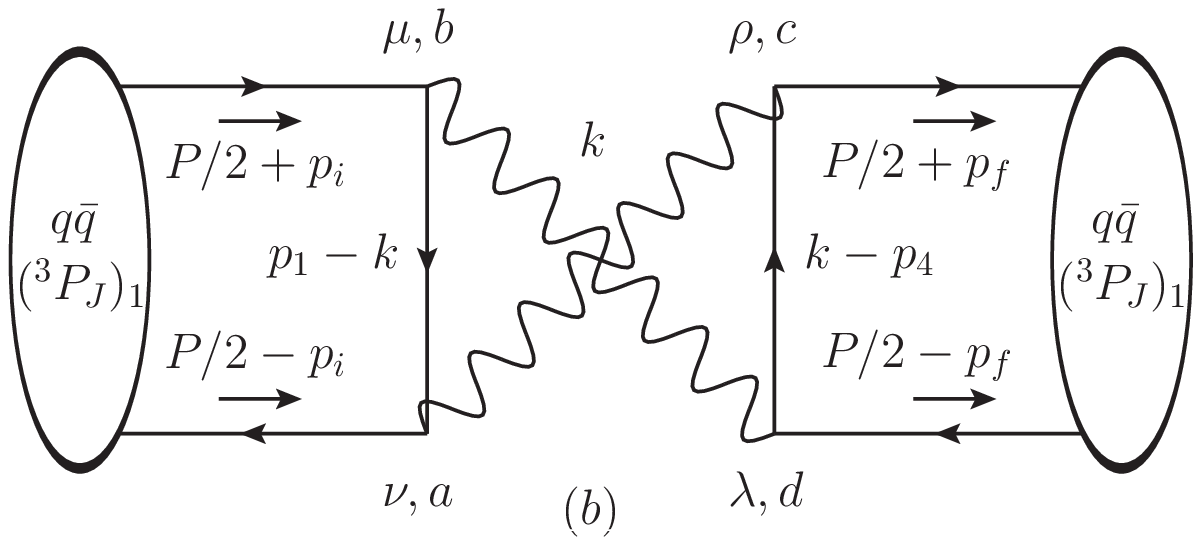}}
\caption{The Feynman diagrams for $q\overline{q}\rightarrow 2g\rightarrow q\overline{q}$ in the $^{3}P_J$ channels in the leading order of $\alpha_s$.}
\label{figure:TGE-project}
\end{figure}

\begin{figure}[htbp]
\center{\epsfxsize 3.3 truein\epsfbox{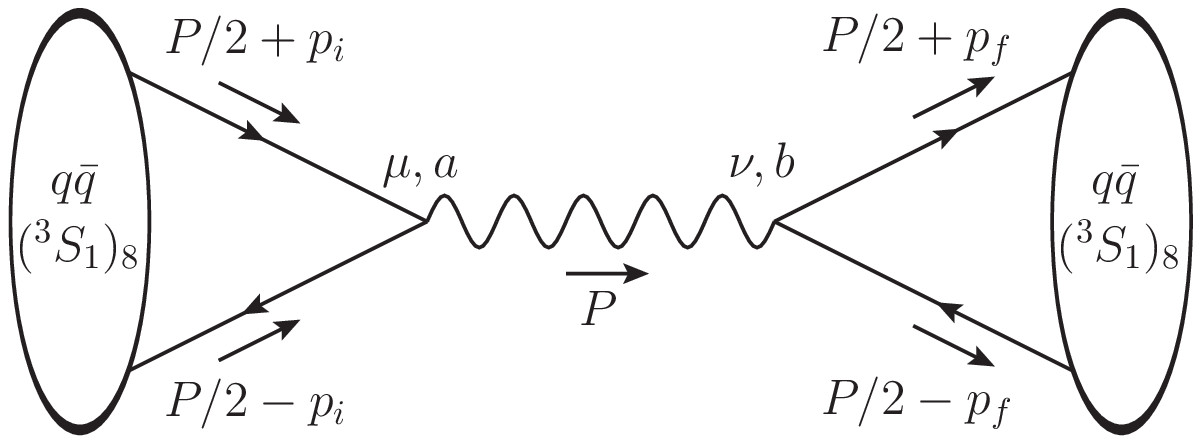}}
\caption{The Feynman diagrams for $q\overline{q}\rightarrow 1g\rightarrow q\overline{q}$ in the $^{3}S_1$ channel in the leading order of $\alpha_s$.}
\label{figure:OGE-project}
\end{figure}

Following the idea of NRQCD, for a heavy quarkonium $H(^3P_J)$ in $J^{++}$ state there are two contributions in the amplitudes of $H(^3P_J)\rightarrow H(^3P_J)$ in the leading order of $v$ (LO-$v$) which can be expressed as
\begin{eqnarray}
\mathcal{M}(H(^3P_J)\rightarrow H(^3P_J)) &= & \mathcal{M}(q\bar{q}(^3P_J)_1\rightarrow q\bar{q}(^3P_J)_1)H_1+\mathcal{M}(q\bar{q}(^3S_1)_8\rightarrow q\bar{q}(^3S_1)_8)H_8,
\end{eqnarray}
where $H_1$ and $H_8$ are some nonperturbative matrix elements, $\mathcal{M}(q\bar{q}(^3P_J)_1\rightarrow q\bar{q}(^3P_J)_1) $ and $\mathcal{M}(q\bar{q}(^3S_1)_8\rightarrow q\bar{q}(^3S_1)_8)$ are the amplitudes with the momenta of the quarks and antiquarks on shell and the indexes 1 and 8 refer to the color signal and color octet states, respectively. The amplitudes at quark level  can be calculated perturbatively. In the perturbation theory, the corresponding Feynman diagrams for the amplitudes of $q\bar{q}(^3P_J)_1\rightarrow 2g\rightarrow q\bar{q}(^3P_J)_1$ are shown in Fig. \ref{figure:TGE-project}, and the the transition $q\bar{q}(^3S_1)_8\rightarrow g\rightarrow q\bar{q}(^3S_1)_8$ are shown in Fig. \ref{figure:OGE-project}.

In the center mass frame, we choose the momenta as follows.
\begin{eqnarray}
p_1 &\triangleq& \frac{1}{2}P+p_i,~~~~p_2 \triangleq \frac{1}{2}P-p_i, \nonumber \\
p_3 &\triangleq& \frac{1}{2}P+p_f,~~~~p_4 \triangleq \frac{1}{2}P-p_f,
\end{eqnarray}
with $p_1^2=p_2^2=p_3^2=p_4^2=m^2$ and $m$ the mass of heavy quark. We can define $P \triangleq (E,0,0,0)$, $p_i\triangleq(0,\textbf{p}_i)$, and $p_f\triangleq(0,\textbf{p}_f)$. For the heavy quark and antiquark pairs we can take $|\textbf{p}_{i,f}|/m$ as small variables, and then can expand the expressions on these small variables. In the on-shell case, we have the relations $|\textbf{p}_i| = |\textbf{p}_f|\triangleq p$ and $E=\sqrt{m^2+p^2}$. This relation means we can not distinguish $|\textbf{p}_i|$ from $|\textbf{p}_f|$ and a non-uniqueness may happen when applying the final expressions to the bound states. There is no such non-uniqueness in the direct calculation of the imaginary parts since the momenta $p_i$ and $p_f$ appear independently after cutting the gluon lines. Fortunately, we can see the first two non-zero orders can be gotten uniquely due to the symmetry which will be discussed in the following.

To project the quark and antiquark pairs to the $^3P_J$ state and the $^3S_1$ state, we use the project matrix in the on-shell case \cite{project-operator-1,project-operator-2} and have the following.
\begin{eqnarray}
\sum \overline{v}(p_2,s_2) T u(p_1,s_1)<\frac{1}{2}s_1;\frac{1}{2}s_2|1s_i> & \triangleq& \textrm{Tr}[T.\Pi_{ini}(s_i)], \nonumber\\
\sum \overline{u}(p_3,s_3) T v(p_4,s_4)<\frac{1}{2}s_3;\frac{1}{2}s_4|1s_f> & \triangleq & \textrm{Tr}[T.\Pi_{fin}(s_f)],
\end{eqnarray}
where the Clebsch-Gordan coefficients are the standard ones as in Ref. \cite{project-operator-2} and the Dirac spinors are normalized as $u^+u=v^+v=1$, whose expressions are written as
\begin{eqnarray}
u(p_1,s_1) &\triangleq&  \frac{\sla{p_}_1+m}{ \sqrt{E_1(E_1+ m)}} \begin{pmatrix} \xi^{s_1}\\0 \end{pmatrix}, \nonumber \\
v(p_2,s_2)&\triangleq& \frac{-\sla{p_}_2+m}{  \sqrt{E_2(E_2+ m)}} \begin{pmatrix} 0\\ \eta^{s_2} \end{pmatrix},
\end{eqnarray}
with $E_{1,2}\triangleq\sqrt{\textbf{p}_{1,2}^2+m^2}=E$, $\xi^{1/2}= (1,0)^T$, $\xi^{-1/2}= (0,1)^T$, $\eta^{1/2}= (0,1)^T$, and $\eta^{-1/2}= (-1,0)^T$. The combination of the above expressions results in the following:
\begin{eqnarray}
\Pi_{i}(s_i) &=&  -\frac{1}{8\sqrt{2} E_i^2 (E_i + m)}
(\sla{p}_1+m)(2E_i+\sla{P})\sla{\epsilon}(s_i)(-\sla{p}_2+m),\nonumber \\
\Pi_{f}(s_f) &=& -\frac{1}{8\sqrt{2} E_f^2 (E_f + m)} (-\sla{p}_4+m)\sla{\epsilon}^{*}(s_f)(2E_s+\sla{P})(\sla{p}_3+m),
\label{eq:project-matirx}
\end{eqnarray}
where $E_{i}=E_{f} \triangleq \sqrt{\textbf{p}_{i,f}^2+m^2}$ and
\begin{eqnarray}
\epsilon^\mu(0) & \triangleq& (0,0,0,1) , \nonumber \\
\epsilon^\mu(\pm1) &\triangleq& (0,\mp1,-i,0)/\sqrt{2}.
\end{eqnarray}
Here the relative sign of $\Pi_{i}(s_i)$ and $\Pi_{f}(s_i)$ is positive which is different from the $^1S_0$ case.

In our calculation, we only calculate the amplitudes in the perturbative QCD and do not go to match them with the corresponding amplitudes in NRQCD in the on-shell region, so we directly try to include the structure of the heavy quarkonia in our calculation.  To include the information of  the $H(^3P_J)$ in the perturbative QCD, we assume the following structure for the $H(^3P_J)$.
\begin{eqnarray}
|H(^3P_J)> &\sim & \phi_1(|\textbf{p}|)\frac{\delta_{ij}}{\sqrt{N_c}}|(q^i\overline{q}^j)_1(^3P_J)>+\phi_0(|\textbf{p}|)\frac{T_a^{ij}}{\sqrt{(N_c^2-1)/2}}|(q^i\overline{q}^j)_8(^3S_1)g^a>
\end{eqnarray}
where the color factors $1/\sqrt{N_c}$ and $1/\sqrt{(N_c^2-1)/2}$ are used to normalized the color parts to 1,  $\phi_{0,1}(|\textbf{p}|)$ refer to the wave functions of the $H(^3P_J)$ in the momentum space in the color single and color octet states, respectively. The relations between the wave functions $\phi_{0,1}$  with the wave functions in the coordinate space are defined as
\begin{eqnarray}
\phi_{l}(|\textbf{p}|)Y_{lm}(\Omega_{\textbf{p}})\triangleq\int d^3\textbf{r} \frac{1}{(2\pi)^3}e^{-i\textbf{p}\cdot\textbf{r}}R_l(|\textbf{r}|)Y_{lm}(\Omega_{\textbf{r}}).
\end{eqnarray}

Using the structure of the $H(^3P_J)$ and the above project matrices, the amplitudes can be expressed as follows.
\begin{eqnarray}
\mathcal{M}(^3P_J) &\triangleq& \mathcal{M}(q\bar{q}(^3P_J)_1\rightarrow q\bar{q}(^3P_J)_1)H_1 \nonumber\\
&=& \int d|\textbf{p}_i|d|\textbf{p}_f| |\textbf{p}_i|^2|\textbf{p}_f|^2\phi_1(|\textbf{p}_f|)\phi_1^*(|\textbf{p}_i|)\overline{G}^{(a+b)}(^3P_J),\nonumber\\
\mathcal{M}(^3S_1)>&\triangleq& \mathcal{M}(q\bar{q}(^3S_1)_8\rightarrow q\bar{q}(^3S_1)_8)H_8\nonumber\\
&=& \int d|\textbf{p}_i|d|\textbf{p}_f| |\textbf{p}_i|^2|\textbf{p}_f|^2\phi_0(|\textbf{p}_f|)\phi_0^*(|\textbf{p}_i|) \overline{G}^{(c)}(^3S_1),
\end{eqnarray}
where $\overline{G}^{(a,b)}(^3P_J)$ and $\overline{G}^{(c)}(^3S_1)$ are expressed as
\begin{eqnarray}
\overline{G}^{(a,b)}(^3P_J)&=&  \sum_{s_i,s_f} <JJ_z|1s_{f};1m_f><JJ_z|1s_{i};1m_i>\int d\Omega_{\textbf{p}_i}d\Omega_{\textbf{p}_f}Y_{1m_i}(\Omega_i)Y^*_{1m_f}(\Omega_f) G^{(a,b)}(s_i,s_f),\nonumber\\
\overline{G}^{(c)}(^3S_1)&=& <1J_z1s_{f}|> <1s_{i}|1J_z> \int d\Omega_{\textbf{p}_i}d\Omega_{\textbf{p}_f}Y_{00}(\Omega_i)Y^*_{00}(\Omega_f) G^{(c)}(s_i,s_f),
\end{eqnarray}
and
\begin{eqnarray}
G^{(a)}(s_i,s_f)&=& -ic_f^{(2g)} \int \frac{d^D k}{(2\pi)^D} \textrm{Tr}[T_1   \Pi_{i}(s_i)] \textrm{Tr}[T_2   \Pi_{f}(s_f)]  D_{\mu\rho}(k)D_{\nu\lambda}(p_1+p_2-k),  \nonumber\\
G^{(b)}(s_i,s_f)&=& -ic_f^{(2g)} \int  \frac{d^D k}{(2\pi)^D} \textrm{Tr}[T_1  \Pi_{i}(s_i)] \textrm{Tr}[T_3   \Pi_{f}(s_f)]  D_{\mu\lambda} (k)D_{\nu\rho} (p_1+p_2-k), \nonumber\\
G^{(c)}(s_i,s_f)&=& -ic_f^{(1g)} \textrm{Tr}[(-ig_s\gamma^{\mu} \Pi_{i}(s_i)] \textrm{Tr}[(-ig_s \gamma^{\nu} \Pi_{f}(s_f)]  D_{\mu\nu} (p_1+p_2), \label{Gabc}
\end{eqnarray}
with $D=4-2\epsilon$ and the color factor $c_f^{(2g)}$
\begin{eqnarray}
c_f^{(2g)}&=&(\frac{\delta_{ij}}{\sqrt{N_c}} T_a^{jm} T_b^{mi}) (\frac{\delta_{i'j'}}{\sqrt{N_c}} T_c^{j'm'} T_d^{m'i'}) \delta_{ad}\delta_{bc} \nonumber \\
&=&\frac{C_A C_F}{2N_c}=\frac{N_c^2-1}{4N_c}=\frac{2}{3},\nonumber \\
c_f^{(1g)}&=&(\frac{T_a^{ij}}{\sqrt{(N_c^2-1)/2}} T_b^{ij}) (\frac{T_c^{i'j'} }{\sqrt{(N_c^2-1)/2}} T_d^{i'j'} )\delta_{bc}=\frac{1}{2},
\end{eqnarray}
the hard kernel $T_i$
\begin{eqnarray}
T_1 &=& (-ig_s \gamma^{\nu}) \cdot S_F(p_1-k) \cdot  (-ig_s \gamma^\mu), \nonumber\\
T_2 &=& (-ig_s \gamma^\rho) \cdot  S_F(p_3-k)  \cdot (-ig_s \gamma^{\lambda}), \nonumber\\
T_3 &=& (-ig_s \gamma^{\rho}) \cdot S_F(k-p_4) \cdot (-ig_s \gamma^{\lambda}),
\end{eqnarray}
and
\begin{eqnarray}
S_F(q) &=& \frac{i(\sla{q}+m)}{q^2-{m}^2+i\varepsilon}, \nonumber \\
D_{\mu\rho}(q) &=& \frac{-i }{q^2+i\varepsilon}(g_{\mu\rho}- \xi \frac{q_\mu q_\rho}{q^2}).
\end{eqnarray}
In the real bound states the values of $|\textbf{p}_i|$ and $|\textbf{p}_f|$ are independent which is different from the on-shell case, so we label $|\textbf{p}_i|$ and $|\textbf{p}_f|$ independently in the above original expressions.

To calculate $G^{(a,b,c)}(s_i,s_f)$, we  use the package Feyncalc \cite{FeynCalc} to do the trace of Dirac matrices in $D$-dimension and then expand the expressions on the variable $p$ to a special order. After the expansion, we use the tensor decomposition to re-expressed the loop integrations and finally use the package FIESTA \cite{FIESTA} to do sector decomposition and then use Mathematica to do the analysic integration.

After the the loop integrations, the form of $G^{(a+b,c)}(s_i,s_f)$ can be expressed as follows.
\textcolor{black}{
\begin{eqnarray}
G^{(a+b,c)}(s_i,s_f)& = &  C^{(a+b,c)}_1 \epsilon(s_i) \cdot \epsilon^*(s_f) +C^{(a+b,c)}_2 \epsilon(s_i) \cdot p_i \epsilon^*(s_f) \cdot p_f+C^{(a+b,c)}_3 \epsilon(s_i) \cdot p_f \epsilon^*(s_f) \cdot p_i\nonumber\\
&&+C^{(a+b,c)}_4 \epsilon(s_i) \cdot p_i \epsilon^*(s_f) \cdot p_i+C^{(a+b,c)}_5 \epsilon(s_i) \cdot p_f \epsilon^*(s_f) \cdot p_f,
\end{eqnarray}}
with
\textcolor{black}{
\begin{eqnarray}
C^{(a+b,c)}_{i}& = & \sum_{n=0}^{3} C^{(a+b,c)}_{in}(p_i^2,p_f^2) (p_i \cdot p_f)^n.
\end{eqnarray}}
After getting the coefficients $C^{(a+b,c)}_{in}$, usually the properties of the integrations of angle and the sums of the spins are independently used to simplify the expressions as in Ref. \cite{project-operator-1}. In our calculation, for simplification we directly calculate the sums of the spins and the integrations of angles together. We define
\begin{eqnarray}
P(J,X,n) &\triangleq& \sum_{s_i,s_f}<JJ_z|1s_{f};1m_f><JJ_z|1s_{i};1m_i> \int d\Omega_{\textbf{p}_i}d\Omega_{\textbf{p}_f}Y_{1m_i}(\Omega_i)Y^*_{1m_f}(\Omega_f)(\hat{p}_i\cdot \hat{p}_f)^n X,\nonumber\\
Q(X,n) &\triangleq&  <1J_z|1s_{f}><1s_{i}|1J_z>\int d\Omega_{\textbf{p}_i}d\Omega_{\textbf{p}_f}Y_{00}(\Omega_i)Y^*_{00}(\Omega_f)(\hat{p}_i\cdot \hat{p}_f)^n X,
\end{eqnarray}
where $X$ are some functions dependent on $\hat{p}_i$, $\hat{p}_f$, $\epsilon(s_i)$ and $\epsilon^*(s_f)$ with $\hat{p}_{i,f}\triangleq p_{i,f}/|\textbf{p}_{i,f}|$, $n=0,1,2$ and $3$, $J=1$, $2$ and $3$, $P(X,n)$ and $Q(J,X,n)$ are not dependent on $J_z$ whose manifest form are directly listed in Appendix A.  Using the expressions of $P(J,X,n)$ and $Q(X,n)$, $\mathcal{M}(^3P_J)$ and $\mathcal{M}(^3S_1)$ can be calculated easily.

\section{\label{sec-3} The analytic results for the asymptotic behavior}

In the practical calculation, we expand the expressions on $p$ to order 6. Since the calculation is taken with the momenta on-shell, the gauge invariance is manifest. The final result can be expressed as
\begin{eqnarray}
\overline{G}^{(a+b)}(^3P_J)\Big{|}_{p} & = & c_f^{(2g)} \alpha_s^2 \pi
\Big [\frac{p^2}{m^4}c_{J,2}+\frac{p^4}{m^6}c_{J,4}+\frac{p^6}{m^8}c_{J,6} + \textrm{higher order} \Big],\nonumber\\
\overline{G}^{(c)}(^3S_1)\Big{|}_{p} & = & c_f^{(1g)} \alpha_s^2 \pi
\Big [\frac{1}{m^2}d_{0}+\frac{p^2}{m^4}d_{2}+\frac{p^4}{m^6}d_{4} + \textrm{higher order} \Big],
\end{eqnarray}
where the subindexes $p$ means to expand the expressions on $p$. For the real bound states, the terms $p^2$ and $p^4$ only receive contributions from the terms $|\textbf{p}_i||\textbf{p}_f|$ and $\frac{1}{2}(|\textbf{p}_i||\textbf{p}_f|^3+|\textbf{p}_i|^3|\textbf{p}_f|)$, respectively, since only they are nonzero. The term $p^6$ receives the contributions both from the terms  $\frac{1}{2}(|\textbf{p}_i||\textbf{p}_f|^5+|\textbf{p}_i|^5|\textbf{p}_f|)$ and $|\textbf{p}_i|^3|\textbf{p}_f|^3$, which results in that one can not distinguish them in a unitary from.

The imaginary parts of $c_{J,i}$ and $d_i$ are expressed as follows.
\begin{align}
 \textrm{Im}[c_{0,2}]=8\pi,~~~~\textrm{Im}[c_{0,4}]&=-\frac{56}{3}\pi,~~~~\textrm{Im}[c_{0,6}]=\frac{1384}{45}\pi, \nonumber \\
\textrm{Im}[c_{1,i}] &= 0,  \nonumber\\
 \textrm{Im}[c_{2,2}] =\frac{32}{15}\pi,~~~~\textrm{Im}[c_{2,4}]&=-\frac{64}{15}\pi,~~~~\textrm{Im}[c_{2,6}]=\frac{51088}{7875}\pi,  \nonumber\\
\textrm{Im}[d_{i}] &=0.
\label{eq:on-shell-Im}
\end{align}
The real parts of $c_{J,i}$ and $d_i$ are expressed as follows.
\begin{eqnarray}
\textrm{Re}[c_{0,2}] &=&-\frac{8}{3}- 16     \log 2 - \frac{64}{9}  C_{IR} ,\nonumber \\
\textrm{Re}[c_{0,4}]&=&\frac{32}{45}+\frac{112}{3}\log2+\frac{832}{45}  C_{IR} ,\nonumber \\
\textrm{Re}[c_{0,6}] &=&\frac{9452}{1575}-\frac{2786}{45}\log2-\frac{17216}{525}   C_{IR},
\label{eq:on-shell-Re-3P0}
\end{eqnarray}
\begin{eqnarray}
\textrm{Re}[c_{1,2}]&=&-\frac{16}{9}- \frac{64}{9}  C_{IR},\nonumber \\
\textrm{Re}[c_{1,4}]&=&\frac{16}{45}+\frac{512}{45}  C_{IR},\nonumber \\
\textrm{Re}[c_{1,6}]&=&\frac{464}{315 }-\frac{22528}{1575} C_{IR},
\label{eq:on-shell-Re-3P1}
\end{eqnarray}
\begin{eqnarray}
\textrm{Re}[c_{2,2}] &=&\frac{32}{15}-\frac{64}{15}\log2- \frac{64}{9}  C_{IR} ,\nonumber \\
\textrm{Re}[c_{2,4}]&=&-\frac{1568}{225}+\frac{128}{15}\log2+\frac{128}{9}  C_{IR},\nonumber \\
\textrm{Re}[c_{2,6}] &=&\frac{14656}{1125}-\frac{102176}{7875}\log2-\frac{11168}{525}  C_{IR}.
\label{eq:on-shell-Re-3P2}
\end{eqnarray}
\begin{eqnarray}
\textrm{Re}[d_{0}] &=&-8\pi,\nonumber \\
\textrm{Re}[d_{2}]&=& 8\pi,\nonumber \\
\textrm{Re}[d_{4}] &=& -\frac{74}{9}\pi.
\label{eq:on-shell-Re-3S1}
\end{eqnarray}
with
\begin{eqnarray}
C_{IR}&=&-\frac{1}{2}\Big(\frac{1}{\epsilon}-\log\frac{m^2}{4 \pi  \mu_{IR}^2}-\gamma_E \Big).
\end{eqnarray}
In the following discussion, we define $\textrm{Re}[c_{J,i}^{fin}]$ as the finite part of $\textrm{Re}[c_{J,i}]$ with the $C_{IR}$ related parts  being subtracted.

Using the above expressions and the quasi potential method, one has the following relation for the corresponding effective potential in the LO-$\alpha_s$.
\begin{eqnarray}
<V_{eff,J}>\triangleq <H(^3P_J)|V_{eff}|H(^3P_J)> &=& -(\mathcal{M}(^3P_J)+\mathcal{M}(^3S_1)).
\end{eqnarray}

In the LO-$v$, the corresponding decay widths of $H(^3P_J)$ to the light hadrons ($l.h$) from the above diagrams which is labeled as $\Gamma(^3P_J\rightarrow l. h)$ are expresses as
\begin{eqnarray}
\Gamma(^3P_J \rightarrow l.h)= -2\textrm{Im}[<V_{eff,J}>] =\frac{3}{4\pi}\alpha_s^2 \textrm{Im}[c_{J,2}] \frac{\big|R_1^{(1)}(0)\big|^2}{m^4},
\label{Eq:Decaywidth-our}
\end{eqnarray}
and the corresponding mass shifts labeled as $\Delta M(^3P_J)$ are expressed as
\begin{eqnarray}
\Delta M(^3P_J)&=&\textrm{Re}[<V_{eff,J}>] \nonumber\\
&=& -\frac{3}{8\pi}\alpha_s^2 \textrm{Re}[c_{J,2}^{fin}] \frac{\big|R_P^{(1)}(0)\big|^2}{m^4}+\pi\alpha_s \Big[\frac{8}{3}\frac{\alpha_s}{\pi^2} C_{IR}  \frac{\big|R_1^{(1)}(0)\big|^2}{m^4}+\frac{1}{4\pi}\frac{\big|R_0(0)\big|^2}{m^2}\Big],
\label{Eq:Delta-m}
\end{eqnarray}
where we have used the relation
\begin{eqnarray}
\int \phi_1(p) p^{2n+3} dp & = &  (-1)^{n}\frac{2n+3}{4\pi}R_1^{(2n+1)}(|\textbf{r}|) \Big |_{|\textbf{r}|=0}, \nonumber\\
\int \phi_0(p) p^{2n+2} dp & = &  (-1)^{n}\frac{1}{4\pi}R_0^{(2n)}(|\textbf{r}|) \Big |_{|\textbf{r}|=0}.
\end{eqnarray}

Eq.(\ref{Eq:Delta-m}) shows when one goes to discuss the mass shifts of the $H(^3P_J)$ states due to the two-gluon annihilation effects in the LO-$\alpha_s$, the color octet contribution should also be considered. The IR divergence in $\mathcal{M}(^3P_J)$ can be absorbed in a unitary way by $\mathcal{M}(^3S_1)$ in the LO-$v$. The absorbed form is unique and same with the case of the one-loop radiative corrections to the decay width $\Gamma(^3P_J \rightarrow l.h)$. This is nature if one goes to match the above results with the corresponding NRQCD coefficients.

In the literature, the decay widths $\Gamma(^3P_J\rightarrow l. h)$ in the LO-$v$ and the NLO-$\alpha_s$  can be expressed as follows \cite{Petrelli1997}.
\begin{eqnarray}
\Gamma(\chi_{0} \rightarrow LH) &=& \frac{4}{3}  \pi \alpha_s^2 H_1\Big[1+\frac{\alpha_s C_0}{\pi}\Big] + n_f \frac{\pi}{3}\alpha_s^2\Big[\frac{16}{27}\frac{\alpha_s}{\pi}H_1\log\frac{m}{\cal E} + H_8 \Big] \nonumber\\
\Gamma(\chi_{2} \rightarrow LH) &=& \frac{16}{45}  \pi \alpha_s^2 H_1 \Big[1+ \frac{\alpha_s}{\pi} C_2 \Big] +n_f \frac{\pi}{3}\alpha_s^2\Big[\frac{16}{27}\frac{\alpha_s}{\pi}H_1\log\frac{m}{\cal E} + H_8 \Big],
\label{Eq:Decaywidth-ref1}
\end{eqnarray}
where $n_f$ is the number of  light quarks, $n_f = 3$ for charmonium and $n_f = 4$ for bottomonium states, $C_0$ and $C_2$ are expressed as
 \begin{eqnarray}
 C_0&=&\frac{4}{3}  (\frac{\pi ^2}{4}-\frac{7}{3} )+3  (\frac{454}{81}-\frac{\pi^2}{144}-\frac{11 \log2}{3})+3  (\frac{2 \log 2}{3}-\frac{16}{27} ),\nonumber\\
 C_2&=&-\frac{16}{3}+3  (\frac{2239}{216}-\frac{337 \pi^2}{384}-2 \log2 )+3 (\frac{2 \log2}{3}-\frac{11}{8} ),
\end{eqnarray}
and $H_1$ is related to the derivative of the wave function through the relation:
\begin{eqnarray}
H_1 = \frac{9}{2\pi} \frac{|R_1^{(1)}|^2}{m^4}[1+O(v^2)].
\label{Eq:H1}
\end{eqnarray}
After the following replacement similarly with that in \cite{NRQCD-ZhaoGuangDa1996}:
\begin{eqnarray}
\log\frac{m}{\cal E} \sim \frac{1}{-2\epsilon}\sim C_{IR},
\end{eqnarray}
we can see that the IR divergences in Eq.(\ref{Eq:Delta-m}) and Eq. (\ref{Eq:Decaywidth-ref1}) are absorbed in the same form.

In the literature, the relation between $H_8$ and the wave function of the high Fock state is not given. From the comparison between Eq. (\ref{Eq:Delta-m}) and Eq. (\ref{Eq:Decaywidth-ref1}) we can get the following relation:
\begin{eqnarray}
H_8  &=& \frac{1}{4\pi} \frac{|R_0(0)|^2}{m^2}.
\label{Eq:H8}
\end{eqnarray}

Combing Eq.(\ref{Eq:Delta-m}) with
Eq. (\ref{Eq:Decaywidth-ref1}), finally one can get
\begin{eqnarray}
\Delta M(^3P_J) &=& -\frac{1}{12}\textrm{Re}[c_{J,2}]\alpha_s^2H_1+\pi\alpha_s \overline{H}_8,
\label{Eq:Delta-m-final}
\end{eqnarray}
with
\begin{eqnarray}
\overline{H}_8&\triangleq&\frac{8}{3}\frac{\alpha_s}{\pi^2} C_{IR}  \frac{\big|R_1^{(1)}(0)\big|^2}{m^4}+\frac{1}{4\pi}\frac{\big|R_0(0)\big|^2}{m^2}\nonumber\\
&\sim&\frac{16}{27}\frac{\alpha_s}{\pi}H_1\log\frac{m}{\cal E} + H_8,
\end{eqnarray}
Eq. (\ref{Eq:Delta-m-final}) can be used to estimate the mass shifts of $H(^3P_J)$ in the LO-$\alpha_s$ and the LO-$v$ due to the two-gluon annihilation effect.

Furthermore, comparing Eq.(\ref{eq:on-shell-Im}) with the decay width $\Gamma(^3P_{J} \rightarrow 2\gamma)$ \cite{Brambilla2006} in NRQCD which expressed as
\begin{eqnarray}
\Gamma (^3P_{0} \rightarrow \gamma \gamma) =&&\frac{6\alpha_{QED}^2 Q^4 \pi}{m^4}\langle ^3P_{0} |\mathcal
O_{\textrm{em}} (^3P_0)| ^3P_{0} \rangle-\frac{14\alpha_{QED}^2 Q^4 \pi}{m^6}\langle ^3P_{0} |\mathcal
P_{\textrm{em}} (^3P_0)| ^3P_{0} \rangle\nonumber\\
&&-\frac{3\alpha_{QED}^2 Q^4 \pi}{m^5}\langle ^3P_{0} |\mathcal
T_{8\,\textrm{em}} (^3P_0)| ^3P_{0} \rangle,\nonumber\\
\Gamma (^3P_2 \rightarrow \gamma \gamma) =&&
\frac{8\alpha_{QED}^2 Q^4 \pi}{5 m^4}\langle ^3P_2 |\mathcal
O_{\textrm{em}} (^3P_2)| ^3P_2 \rangle- \frac{16\alpha_{QED}^2 Q^4 \pi}{5 m^6}\langle ^3P_2 |\mathcal
P_{\textrm{em}} (^3P_2)| ^3P_2 \rangle,
\label{eq:NQRCD-2gamma}
\end{eqnarray}
one can also see the first and second terms in Eqs. (\ref{eq:on-shell-Im}) are just same with the coefficients of the first and second terms of Eq. (\ref{eq:NQRCD-2gamma}) except for a global different color factor. The coefficients $\textrm{Im}[c_{J,6}]$ should be same with the sum of the corresponding coefficients in NRQCD in order $v^6$.

\section{Numerical result and conclusion}
The main results of our calculation are the expressions of the coefficients $c_{J,i}$ and the mass shifts $\Delta M(^3P_J)$. In the LO-$v$ and the LO-$\alpha_s$, if one assumes that the contribution from the $\overline{H}_8$ related term is small, then the ratios between the mass shifts and the decay widths can be expressed as
\begin{eqnarray}
\frac{\Delta M(^3P_0)}{\Gamma(^3P_0)}&=&\frac{1+6\log2}{6\pi}\approx0.27,\nonumber\\
\frac{\Delta M(^3P_2)}{\Gamma(^3P_2)}&=&\frac{2\log2-1}{2\pi}\approx0.06,\nonumber\\
\frac{\Delta M_{^1S_0}}{\Gamma(^1S_0)}&=&\frac{\log2-1}{\pi}\approx-0.098,
\end{eqnarray}
where the similar result for the $^1S_0$ state is also presented. We can see that the ratio for the $^3P_0$ state is much larger than the ratios for the $^3P_2$ and $^1S_0$ states and the ratios are positive for the $^3P_J$ states and negative for the $^1S_0$ state.

Furthermore one can extract the parameters $H_1$ and $\overline{H}_8$ from the experimental data by Eq. (\ref{Eq:Decaywidth-ref1}) in the NLO-$\alpha_s$ and the LO-$v$, then one can use the extracted parameters to estimate the mass shifts $\Delta M(^3P_J)$ using Eq. (\ref{Eq:Delta-m-final}) in the LO-$\alpha_s$ and the LO-$v$. The corresponding numerical results of $\Delta M(^3P_J)$ for $\chi_{cJ}$  are listed in Tab. \ref{Table-mass-shift} where the experimental data are taken from Ref. \cite{PDG2018}. The similar estimation can be applied to the bottomonium. Comparing these numerical results with the corresponding results of $\eta_c$ \cite{CaoHuiYun2018}, we can find that the mass shifts of $\chi_{cJ}$ are very different. These properties mean the corrections to different states can not be subtracted or hidden in a unified way. Combing the numerical results, one can get $\Delta M(^3P_0)-\Delta M(^1S_0)\approx 9.0 \sim 8.6 $ MeV with $\alpha_s\approx 0.25\sim0.35$, correspondingly. This numerical result suggests that the annihilation effects should be considered seriously when try to understand the spectrum of heavy quarkonia precisely, especially when some decay channels with large decay widths are opened.
\begin{table}[hbtp]
\centering
\begin{tabular}{|p{60pt}<{\centering}|p{60pt}<{\centering}|p{1pt}<{\centering}|p{60pt}<{\centering}|p{60pt}<{\centering}|p{90pt}<{\centering}|}
\hline
 &$\Gamma^{Ex}_{l.h}$(MeV)&& $H_1$(MeV)& $\overline{H}_8$(MeV)& $\Delta M(^3P_J)$(MeV)\\
\cline{1-2}\cline{4-6}
$\chi_{c0}(1P)$&  $10.8$&  &\multirow{3}{*}{$69.8\sim29.3$}& \multirow{3}{*}{$1.18\sim1.21$} &$5.92\sim5.45$\\
\cline{1-2}\cline{6-6}
$\chi_{c1}(1P)$&- &  & &   &$1.57\sim1.86$ \\
\cline{1-2}\cline{6-6}	
$\chi_{c1}(2P)$&$1.6$ & &  &   &$1.23\sim1.58$\\
\hline
\end{tabular}
\caption{The numerical results for $\Delta M(^3P_J)$ which refer to the mass shifts of $\chi_{cJ}$ in the leading order. The experimental decay widths are taken from Ref. \cite{PDG2018}, the values of $H_1$ and $\overline{H}_8$ are extracted by using Eq.(\ref{Eq:Decaywidth-ref1}) with $\alpha_s$ taking as $0.25\sim0.35$, correspondingly.}
\label{Table-mass-shift}
\end{table}

Another interesting property  is that although the decay width of the $^3P_1$ states to two-gluon intermediated state is zero, the corresponding mass shift is nonzero.

In summary, the real part of the nonrelativistic asymptotic behavior of the amplitudes of $q\overline{q}\rightarrow2g\rightarrow q\overline{q}$ in the $^3P_J$ channels is discussed in the LO-$\alpha_s$. By expanding the expressions on the three-momentum of quarks and antiquarks, the expressions are calculated to order 6. The imaginary part of the first 2 terms of our results are the same with those given in the references. The real part of our results can be used to estimate the mass shifts of the $^3P_J$ heavy quarkonia due to the two-gluon annihilation effect. In the LO-$\alpha_s$ and the LO-$v$, we get the following properties: (1) the mass shifts of the $^3P_J$ states are positive which are differen form the $^1S_0$ case where the mass shifts are negative; (2) the mass shifts of the $^3P_1$ states are nonzero although their decay widthes are zero; (3) the numerical estimation shows the contributions to the mass shifts of $\chi_{c0,c1,c2}$ are about $1.23\sim1.58$ MeV, $1.57\sim1.86$ MeV and $5.92\sim5.45$ MeV when taking $\alpha_s\approx 0.25\sim0.35$.

\section{Acknowledgments}
The author Hai-Qing Zhou would like to thank Wen-Long Sang, Zhi-Yong Zhou and Dian-Yong Chen for their kind and helpful discussions. This work is supported by the  National Natural Science Foundations of China under Grant No. 11375044.

\section{Appendix A}
In this Appendix, the manifest expressions for $P(J,X,n)$ and $Q(X,n)$ are listed. From the definition of $P(J,X,n)$ and $Q(X,n)$ which are expressed as
\begin{eqnarray}
P(J,X,n) &\triangleq& \sum_{s_i,s_f}<1s_{i};1m_i|JJ_z> <1s_{f};1m_f|JJ_z>\int d\Omega_{\textbf{p}_i}d\Omega_{\textbf{p}_f}Y_{1m_i}(\Omega_i)Y^*_{1m_f}(\Omega_f)(p_i\cdot p_f)^n X,\nonumber\\
Q(X,n) &\triangleq&  <1J_z|1s_{f}><1s_{i}|1J_z>\int d\Omega_{\textbf{p}_i}d\Omega_{\textbf{p}_f}Y_{00}(\Omega_i)Y^*_{00}(\Omega_f)(\hat{p}_i\cdot \hat{p}_f)^n X,
\end{eqnarray}
with $0\leq n\leq 3$, we have
\begin{eqnarray}
&&P(J,\epsilon(s_i) \cdot \epsilon^*(s_f),1)=\frac{4\pi}{3},~~~~P(J,\epsilon(s_i) \cdot \epsilon^*(s_f),3) =  \frac{4\pi}{5},
\end{eqnarray}
\begin{eqnarray}
&&P(0,\epsilon(s_i) \cdot \hat{p}_i \epsilon^*(s_f) \cdot \hat{p}_f,0)=4\pi,~~~~P(0,\epsilon(s_i) \cdot p_i \epsilon^*(s_f) \cdot p_f,2) = \frac{4\pi}{3},\nonumber\\
&&P(2,\epsilon(s_i) \cdot \hat{p}_i \epsilon^*(s_f) \cdot \hat{p}_f,2)=\frac{16\pi}{75},
\end{eqnarray}
\begin{eqnarray}
&&P(0,\epsilon(s_i) \cdot \hat{p}_f \epsilon^*(s_f) \cdot \hat{p}_i,0)=\frac{4\pi}{3},~~~~P(0,\epsilon(s_i) \cdot \hat{p}_f \epsilon^*(s_f) \cdot \hat{p}_i,2) = \frac{4\pi}{5},\nonumber\\
&&P(1,\epsilon(s_i) \cdot \hat{p}_f \epsilon^*(s_f) \cdot \hat{p}_i,0)=-\frac{4\pi}{3},~~~~P(1,\epsilon(s_i) \cdot \hat{p}_f \epsilon^*(s_f) \cdot \hat{p}_i,2) = -\frac{4\pi}{15},\nonumber\\
&&P(2,\epsilon(s_i) \cdot \hat{p}_f \epsilon^*(s_f) \cdot \hat{p}_i,0)=\frac{4\pi}{3},~~~~P(0,\epsilon(s_i) \cdot \hat{p}_f \epsilon^*(s_f) \cdot \hat{p}_i,2) = \frac{12\pi}{25},
\end{eqnarray}
\begin{eqnarray}
&&P(0,\epsilon(s_i) \cdot \hat{p}_i \epsilon^*(s_f) \cdot \hat{p}_i,1)=-\frac{4\pi}{3},~~~~P(0,\epsilon(s_i) \cdot \hat{p}_i \epsilon^*(s_f) \cdot \hat{p}_i,3) = -\frac{4\pi}{5},\nonumber\\
&&P(2,\epsilon(s_i) \cdot \hat{p}_i \epsilon^*(s_f) \cdot \hat{p}_i,1)=-\frac{8\pi}{15},~~~~P(2,\epsilon(s_i) \cdot \hat{p}_i \epsilon^*(s_f) \cdot \hat{p}_i,3) = -\frac{8\pi}{25},
\end{eqnarray}
\begin{eqnarray}
&&P(0,\epsilon(s_i) \cdot \hat{p}_f \epsilon^*(s_f) \cdot \hat{p}_f,1)=-\frac{4\pi}{3},~~~~P(0,\epsilon(s_i) \cdot \hat{p}_f \epsilon^*(s_f) \cdot \hat{p}_f,3) = -\frac{4\pi}{5},\nonumber\\
&&P(2,\epsilon(s_i) \cdot \hat{p}_f \epsilon^*(s_f) \cdot \hat{p}_f,1)=-\frac{8\pi}{15},~~~~P(2,\epsilon(s_i) \cdot \hat{p}_f \epsilon^*(s_f) \cdot \hat{p}_f,3) = -\frac{8\pi}{25},
\end{eqnarray}
and the results for other $(J,n)$ are zero,
\begin{eqnarray}
Q(\epsilon(s_i) \cdot \epsilon^*(s_f),0)&=&-4\pi,~~~~Q(\epsilon(s_i) \cdot \epsilon^*(s_f),2)=-\frac{4\pi}{3}\nonumber \\
Q( \epsilon(s_i) \cdot \hat{p}_i \epsilon^*(s_f) \cdot \hat{p}_f,1)&=&-\frac{4\pi}{9}~~~~Q( \epsilon(s_i) \cdot \hat{p}_i \epsilon^*(s_f) \cdot \hat{p}_f,3)=-\frac{4\pi}{15},\nonumber\\
Q( \epsilon(s_i) \cdot \hat{p}_f \epsilon^*(s_f) \cdot \hat{p}_i,1)&=&-\frac{4\pi}{9},~~~~Q( \epsilon(s_i) \cdot \hat{p}_f \epsilon^*(s_f) \cdot \hat{p}_i,3)=-\frac{4\pi}{15}\nonumber\\
Q( \epsilon(s_i) \cdot \hat{p}_i \epsilon^*(s_f) \cdot \hat{p}_i,1)&=&-\frac{4\pi}{9},~~~~Q( \epsilon(s_i) \cdot \hat{p}_i \epsilon^*(s_f) \cdot \hat{p}_i,3)=-\frac{4\pi}{15}\nonumber\\
Q( \epsilon(s_i) \cdot \hat{p}_i \epsilon^*(s_f) \cdot \hat{p}_i,1)&=&-\frac{4\pi}{9},~~~~Q( \epsilon(s_i) \cdot \hat{p}_f \epsilon^*(s_f) \cdot \hat{p}_f,3)=-\frac{4\pi}{15}
\end{eqnarray}
and the results for other $n$ are zero.

\end{document}